\documentstyle[12pt]{article}

\def\bar{\overline}
\def\hat{\widehat}
\def\*{\star}
\def\({\left(}          
\def\){\right)}         
\def\[{\left[}          
\def\]{\right]}

\def\frac#1#2{{#1 \over #2}}            
\def\inv#1{{1 \over #1}}
\def\half{{1 \over 2}}
\def\d{\partial}

\def\vev#1{\langle #1 \rangle}
\def\ket#1{ | #1 \rangle}
\def\bra#1{ \langle #1 |}

\def\2pi{\hbox{$2\pi i$}}

\def\dsl{\raise.15ex\hbox{/}\kern-.57em\partial}
\def\Dsl{\,\raise.15ex\hbox{/}\mkern-.13.5mu D}
\def\th{\theta}

        \def\La{\Lambda}

\def\sig{\sigma}        

              \def\CC{{\cal C}}
\def\CD{{\cal D}}

\def\debut{ \begin{eqnarray} }
\def\fin{ \end{eqnarray} }
\def\non{ \nonumber }
\begin{document}

\rightline{SPhT-92-062; LPTHE-92-20}
\vskip 1cm
\centerline{\LARGE From Form Factors to Correlation Functions:}
\bigskip
\centerline{\LARGE The Ising Model.}
\vskip 1cm
\centerline{\large Olivier Babelon }
\centerline{Laboratoire de Physique Th\'eorique et Hautes
Energies \footnote[1]{\it Laboratoire associ\'e au CNRS.}}
\centerline{
 Universit\'e Pierre et Marie Curie, Tour 16 1$^{er}$
\'etage, 4
place Jussieu}
\centerline{75252 Paris cedex 05-France}
\vskip1cm
 \centerline{\large  Denis Bernard }
 \centerline{Service de Physique Th\'eorique de Saclay
\footnote[2]{\it Laboratoire de la Direction des Sciences de la
Mati\`ere du Commisariat \`a l'Energie Atomique.}}
\centerline{F-91191, Gif-sur-Yvette, France.}
 \vskip2cm
Abstract.\\
Using exact expressions for the Ising form factors, we give
a new very simple proof that the spin-spin and
disorder-disorder correlation functions are
governed by the Painlev\'e III non linear differential equation.
We also show that the generating function of the correlation
functions of the descendents of the spin and disorder operators
is a $N$-soliton, $N\to\infty$, $\tau$-function of the sinh-Gordon
hierarchy.  We discuss a relation of our approach to isomonodromy
deformation problems, as well as further possible generalizations.

\vfill
\newpage

\section{Introduction.}
In the scaling limit, the Ising model is described by a free
Majorana fermion $\psi (x)$ with mass $m\equiv (T-T_C)$.
However, the order and disorder operators $\sigma (x)$ and $\mu (x)$
are non local in terms of these free fermions.
This shows up in the very striking result by T.T. Wu, B.M. McCoy,
C.A. Tracy and E. Barouch \cite{Coy76} for the correlation functions
of $\sigma(x)$ and $\mu(x)$. Let $r$ be the radial distance, then~:
\debut
\pmatrix{<\mu(r) \mu(0)> \cr <\sigma(r) \sigma(0)> }
 = \pmatrix{\cosh\half\chi (s)\cr \sinh\half\chi (s)}\
\exp\({-{1\over 4}\int^\infty_s du\ u
\Big[ (\frac{d\chi}{du})^2 - \sinh^2\chi \Big]\ }\). \label{IVc}
\fin
where $s$ is the scaling variable $s=\frac{mr}{2}$.
$\chi$ is a solution of the radial sinh-Gordon equation~:
\debut
\frac{d^2\chi}{ds^2} + \inv{s}\frac{d\chi}{ds} = 2\sinh(2\chi) \non
\fin
Setting $\eta = e^{-\chi}$, this is equivalent to the Painlev\'e III
equation:
\begin{eqnarray}
{\eta'' \over \eta } = \left( {\eta'\over \eta }\right)^2 - {1\over s}
 \left( {\eta'\over \eta }\right)+ \eta^2 - {1\over \eta^2} \nonumber
\end{eqnarray}
We present a new (as far as we know) very simple proof of it,
based on the analysis of form factors.

More generally, to compute exact correlation functions in 2D integrable
quantum field theories
remains a challenge despite recent progresses \cite{Its90}. These
progresses
are based on a fine analysis of the algebraic Bethe anstaz.
On the other side, it has recently been argued that the two-dimensional
integrable QFT can be solved by only using their infinite quantum
group symmetries \cite{Be90,Smi91}. The simplest examples of such symmetries
are provided by
the Yangian symmetries of massive current algebras or by the quantum
$sl(2)$ loop symmetry of the sine-Gordon model. These symmetries act
more
simply on the asymptotic states and on the momentum variables.
Therefore,
they lead more directly to algebraic methods for computing field form
factors. The next question which naturally arises in this approach
consists in knowing whether it is possible to reconstruct and to
characterize, e.g. by differential equations, the correlation functions
from the knowledge of the form factors in the spirit of the
boostrap program \cite{ZaZa}. In the following, we show by
examining the simplest possible model, namely the Ising model, that
it is not hopeless to expect a positive answer to this question.

\section{Ising Form Factors.}

As is well known \cite{Isi},
the Ising model is a $Z_2$-invariant lattice model
with nearest neighbour interactions between $Z_2$ spins. It undergoes
a second order phase transition at some critical temperature $T_C$.
In the scaling limit near its critical point it is equivalent
to a free Majorana fermion with mass $m\sim (T-T_c)$.
%In the following we set $m=1$.
It possesses an infinite number of local integrals of motion
in involution which have odd spins. The asymptotic states are the
free fermions and therefore the scattering is trivial~:
\begin{eqnarray}
S=-1
\end{eqnarray}
Among the fields of primary interest are the spin field $\sig$, its
``dual" field called the disorder field $\mu$, and the fermion $\psi$.
We refer to \cite{Ka71} for their lattice definition.

 Form factors are matrix elements of field operators. They satisfy
algebraic relations, usually called form factor axioms
\cite{KW78,Smi}, which depend on
the locality of the fields and on the sectors to which the fields
belong. Due to its $Z_2$ symmetry, the sectors of the Ising model
are labelled by couples of indices $[a;b]$ with $a,\ b = 0,\ 1$.
In each sector $[a;b]$, the first index $a$ refers to a representation
of the group $Z_2$ whereas the second index $b$ refers to an element
in $Z_2$.  The fields $\Phi_{[a;b]}(x,t)$ in sectors  $[a;b]$
satisfy the following equal-time braiding relations~
\footnote{ More generally, in a statistical model invariant under a
finite group $G$, one naturally defines spin fields $\sig^\rho$, which
take values in some representation $\rho$ of $G$, disorder fields
$\mu_g$, where the label $g$ are elements of $G$, and parafermions
$\psi^\rho_g$, which are defined by the operator product expansion
$\psi^\rho_g\sim \mu_g\sig^\rho$. An element $h\in G$ acts on the
parafermions
by $\psi^\rho_g \to \rho(h)\psi^\rho_g$.
The parafermions $\psi^\rho_g$ span the sectors $[\rho;g]$ which form
representations of the quantum double $\CD(G)$ of functions on $G$.
The braiding relations of the parafermions are given by
the universal $R$-matrix of $\CD(G)$:
\begin{eqnarray}
(\psi^{\rho_1}_{g_1}(x_1)\otimes 1)(1\otimes\psi^{\rho_2}_{g_2}(x_2))=
(1\otimes \rho_2(g_1)\psi^{\rho_2}_{g_1g_2g_1^{-1}}(x_2))
(\psi^{\rho_1}_{g_1}(x_1)\otimes 1)\quad;\quad x_1>x_2 \nonumber
\end{eqnarray}
The Ising model is a particular case of that. }:
\debut
\Phi_{[a_1;b_1]}(x_1;t)\ \Phi_{[a_2;b_2]}(x_2;t)\ =\
(-1)^{a_2b_1}\ \Phi_{[a_2;b_2]}(x_2;t)\ \Phi_{[a_1;b_1]}(x_1;t)
\quad;\quad x_1>x_2 \non
\fin
The identity operator is in the sector $[0;0]$, the spin field $\sig$
in $[1;0]$, the disorder operator $\mu$ in $[0;1]$, and the fermion
$\psi$, which is defined by the operator product expansion
$\Psi\sim \mu\sig$, is in the sector $[1;1]$. In particular,
the asymptotic particles, which are the fermions, are in the
sector $[1;1]$.

 In a more algebraic way, the couples $[a;b]$ label the
irreducible representations of the quantum double $\CD(Z_2)$ of
functions
on the group $Z_2$.  In $\CD(Z_2)$, the universal $R$-matrix acts on
two
representations $[a_1;b_1]$ and $[a_2;b_2]$ as $R=(-1)^{a_2b_1}$.
The braiding matrices are therefore given by the universal $R$-matrix.
Moreover, for the representation $[1;1]$ corresponding to the fermions
we find $R=-1$. We therefore have the relation (which is a complicated
way to write
a simple result):
\begin{eqnarray}
S=R_{universal} \non
\end{eqnarray}
As expected from the general arguments \cite{Smi91,BeLe}.

Let us denote by $\ket{\th_1,\cdots,\th_n}$ the $n$-particle states
with energies $E_i=m\cosh\th_i$ and momenta $P_i=m\sinh\th_i$.
The form factors for a field $\Phi_{[a;b]}$ in a sector $[a;b]$
are defined by~:
\debut
F^{(n)}_{[a;b]}(\th_1,\cdots,\th_n)\ \equiv\
\bra{0}\Phi_{[a;b]}(0)\ket{\th_1,\cdots,\th_n} \label{IIIa}
\fin
By crossing symmetry, all the matrix elements of $\Phi_{[a;b]}$
are determined by $F^{(n)}_{[a;b]}$~:
\debut
\bra{\th_1,\cdots,\th_p}\Phi_{[a;b]}(0)
\ket{\th_{p+1},\cdots,\th_n} =
F^{(n)}_{[a;b]}(\th_1-i\pi,\cdots,\th_p-i\pi,
\th_{p+1},\cdots,\th_n)\non
\fin
By $Z_2$-symmetry, the form factors $F^{(n)}_{[a;b]}$ are non-vanishing
only for $a\equiv n\ [mod~2]$.

The form factors axioms are usually written for local fields in the
identity sector \cite{KW78,Smi}. Their generalizations for non-local
fields
in the other sectors is easy to find. In ref \cite{Smi91,BeLe}, the
generalized
axioms were derived using quantum group symmetries.
In the case of the Ising model, since all the sector are described by
the $Z_2\times Z_2$ indices $[a;b]$ and since the $S$-matrix is
$S=-1$, they reduce to~:
\debut
F^{(n)}_{[a;b]}(\th_2,\th_1,\th_3,\cdots,\th_n) &=&
-F^{(n)}_{[a;b]}(\th_1,\th_2,\th_3,\cdots,\th_n) \non\\
F^{(n)}_{[a;b]}(\th_1-i2\pi,\th_2,\cdots,\th_n) &=&
(-1)^{(1+a)(1+b)}\ F^{(n)}_{[a;b]}(\th_1,\th_2,\cdots,\th_n)
\label{IIIb}\\
{\rm Res}_{\th_1=\th_2-i\pi} \
F^{(n)}_{[a;b]}(\th_1,\th_2,\th_3,\cdots,\th_n) &=&
\Bigl[(-1)^{a+b}-1\Bigr] \
F^{(n-2)}_{[a;b]}(\th_3,\cdots,\th_n) \non
\fin
Here, we have used that $a\equiv n\ [mod~2]$. Because there is no bound
state,
the form factors have no other pole in the physical strip,
$0\leq {\rm Im}\th\leq \pi$.

In the sectors $[a;b]$ with $a+b\equiv 0\ [mod~2]$, the form factors
have no pole at all in the physical strip. They correspond to those
of fields in the sectors of the identity and of the fermion.
In the same way as in \cite{CM90}, the set of solutions to the
equations (\ref{IIIb}) with $a+b\equiv 0 \ [mod~2]$ are found
to be in correspondence with the fields in the Neveu-Schwarz
sector of the conformal Ising theory.

The sectors with $a+b\equiv 1\ [mod~2]$ are those of the spin and
disorder
fields. There, the form factors have a simple pole in the physical
strip. This pole corresponds to the particle-antiparticle scattering.
The minimal solution to eqs. (\ref{IIIb}) which possesses no zeroes in
the physical strip is given by \cite{KW78,Sch78,CM90}~:
\debut
F^{(n)}_{min}(\th_1,\cdots,\th_n)\ =\ \prod_{i<j}
\tanh\({\frac{\th_i-\th_j}{2} }\).\label{IIIc}
\fin
These minimal form factors with $n$ odd are identified as those of the
spin field $\sig$, and with $n$ even as those of the disorder
field $\mu$. Eqs. (\ref{IIIb}) determine the minimal solution
up to a multiplicative constant; we choose this constant to be one
by a normalization convention; in particular $\vev{\mu}=1$.

As proved in \cite{KW78}, all the  solutions $F^{(n)}_{[a;b]}$
to eqs. (\ref{IIIb}) can
be written as product of the minimal solution times an auxiliary
function, denoted $P^{(n)}(\th_1,\cdots,\th_n)$, which has
no pole in the physical strip~:
\debut
F^{(n)}_{[a;b]}(\th_1,\cdots,\th_n)\ =\
P^{(n)}(\th_1,\cdots,\th_n) F^{(n)}_{min}(\th_1,\cdots,\th_n)  \non
\fin
Moreover, from eqs. (\ref{IIIb}), it follows that the functions
$P^{(n)}$ satisfy~:
\debut
P^{(n)}(\th_2,\th_1,\th_3,\cdots,\th_n) &=&
P^{(n)}(\th_1,\th_2,\th_3,\cdots,\th_n) \non\\
P^{(n)}(\th_1-i2\pi,\th_2,\cdots,\th_n) &=&
P^{(n)}(\th_1,\th_2,\cdots,\th_n) \label{IIId}\\
P^{(n)}(\th_2+i\pi,\th_2,\th_3,\cdots,\th_n) &=&
P^{(n-2)}(\th_3,\cdots,\th_n) \non
\fin
In other words, $P^{(n)}$ are symmetric and periodic functions
of $\th_1,\cdots,\th_n$ and subject to the last constraints in eq.
(\ref{IIId}).
It is easy to check that an infinite set of solutions to
eqs. (\ref{IIId}) is given by~:
\debut
P^{(n)}_{\{s_1,\cdots,s_M\}}(\th_j)_{j=1,\cdots,n} =
Q^{(n)}_{s_1}(\th_j)\cdots Q^{(n)}_{s_M}(\th_j)\non
\fin
where $s_1,\cdots,s_M$, which label the solutions, are
positive or negative odd integers. The functions $Q_s$ are~:
\debut
Q^{(n)}_s(\th_1,\cdots,\th_n)=-\frac{m}{2} \sum_{j=1}^n\ e^{s\th_j} \non
\fin
As shown in \cite{CM90}, these solutions form a complete set of
independent solutions to eqs. (\ref{IIId}), and are in
one-to-one correspondence with the fields in the Ramond
sector of the Ising conformal theory. As a consequence, the
form factors,
\debut
F^{(2n+1)}_{[1;0],\{s_1,\cdots,s_M\}}(\th_j)_{j=1,\cdots,2n+1} &=&
P^{(2n+1)}_{\{s_1,\cdots,s_M\}}(\th_j) F^{(2n+1)}_{min}(\th_j)  \non \\
F^{(2n)}_{[0;1],\{t_1,\cdots,t_Q\}}(\th_j)_{j=1,\cdots,2n} &=&
P^{(2n)}_{\{t_1,\cdots,t_Q\}}(\th_j) F^{(2n)}_{min}(\th_j) \label{IIIe}
\fin
are identified with the form factors of the descendents
$\sig_{s_1,\cdots,s_M}$ and $\mu_{t_1,\cdots,t_Q}$
of the spin field $\sig$ and disorder field $\mu$.

\section{Ising correlation functions.}

We will only consider the two-point correlation functions. If
$F^{(n)}_1(\th_1,\cdots,\th_n)$ and
$F^{(n)}_2(\th_1,\cdots,\th_n)$ are the form factors of two scalar
fields $\Phi_1(x,t)$ and $\Phi_2(x,t)$, the Euclidean correlation
function is then~:
\debut
\vev{\Phi_1(x,t)\ \Phi_2(0,0)} = \sum_{n=0}^\infty\
\inv{n!}\int_{-\infty}^{+\infty}\ \prod_{j=1}^n
\({\frac{d\th_j}{2\pi} e^{-mr\cosh\th_j}}\)\
F^{(n)}_1(\th_1,\cdots,\th_n) F^{(n)}_2(\th_1,\cdots,\th_n)
\label{IVa}
\fin
where $r$ is the radial distance, $r^2=x^2+t^2$.

Let us first consider the spin-spin and disorder-disorder correlation
functions. Define~:
\debut
C_\pm(r)\ = \vev{\mu(r)\mu(0)} \pm \vev{\sig(r)\sig(0)} \non
\fin
  From the formula for the form factors, we have~:
\debut
C_\pm(r)\ =\ \sum_{n=0}^\infty\ \frac{(\pm)^n}{n!}
\int_{-\infty}^{+\infty}\
\prod_{j=1}^n \({\frac{d\th_j}{2\pi} e^{-mr\cosh\th_j}}\)\
\prod_{i<j} \tanh^2\({\frac{\th_i-\th_j}{2} }\) \label{IVb}
\fin
We will study the sum of this series in the following section.

Consider now the correlation function of the descendents of the spin
and disorder fields. Set~:
\begin{eqnarray}
C^{\{s_p\};\{t_q\}}_\pm(x,t)\ =
\vev{\mu^{\{s_p\}}(x,t)\mu^{\{t_q\}}(0)} \pm
\vev{\sig^{\{s_p\}}(x,t)\sig^{\{t_q\}}(0)} \non
\end{eqnarray}
Since the operators $\sig^{\{s_p\}}$ and $\mu^{\{t_q\}}$
have non-trivial Lorentz spins, the correlation functions
do not only depend on the radial distance but on $x$ and $t$.
We introduce the Euclidean coordinates $z_\pm=x\pm it$. From
the expression of the form factors, we have~:
\begin{eqnarray}
C^{\{s_p\};\{t_q\}}_\pm(x,t)\ =
 \sum_{n=0}^\infty\ \frac{(\pm)^n}{n!}
\int_{-\infty}^{+\infty}\
\prod_{j=1}^n \({\frac{d\th_j}{2\pi}
e^{-\frac{m}{2}(z_+e^{\th_j}+z_- e^{-\th_j})} }\)\
\prod_{p,q} Q_{s_p}(\th_j) Q_{t_q}(\th_j)
\prod_{i<j} \tanh^2\({\frac{\th_i-\th_j}{2} }\) \non
\end{eqnarray}
Here $\{s_p\}$ refers to multi-indices
$\{s_{p_1},\cdots,s_{p_M}\}$ and similarly
$\{t_q\}=\{t_{q_1},\cdots,t_{q_K}\}$.
Let us introduce a generating function $\CC^\infty_\pm(z_s)$
for these correlation functions. It
depends on an infinite number of variables $z_s$ with
$s=\pm1,\pm3,\cdots$; we denote $z_{\pm1}$ by $z_\pm$.
It is specified by~:
\debut
C^{\{s_p\};\{t_q\}}_\pm(x,t) = \prod_{p,q} \frac{\d}{\d z_{s_p}}
\frac{\d}{\d z_{t_q}}\ \CC^\infty_\pm(z_s)
\Big\vert_{ {z_s=0;~~  |s|\geq 3 }}
\non
\fin
 From the exact expressions (\ref{IIIe}) of the form factors
of the descendents, we deduce~:
\debut
\CC^\infty_\pm(z_s) =
\sum_{n=0}^\infty\ \frac{(\pm)^n}{n!}
\int_{-\infty}^{+\infty}\
\prod_{j=1}^n \({\frac{d\th_j}{2\pi} X(\th_j|z_s) }\)\
\prod_{i<j} \tanh^2\({\frac{\th_i-\th_j}{2} }\) \label{IVe}
\fin
where the ``potential" $X(\th|z_s)$ are given by~:
\debut
X(\th|z_s)=\exp\({ -\frac{m}{2}\sum_{s=\pm1,\pm3,\cdots}  z_se^{s\th} }\)
\non
\fin
In the next section, we show that the generating function
$\CC_\pm^\infty(z_s)$ can also be expressed in terms of a solution
of the sinh-Gordon equation.
More precisely we will show that it is a $\tau$-function
for the affine sinh-Gordon hierarchy.

\section{Proof of the differential equations.}

The differential equation for the Ising correlation
functions were originally proved by looking at the scaling
limit of the lattice Ising model \cite{Coy76}. See also \cite{Pe80}.
Another approach was developped by the
Kyoto school which is based on the study of isomonodromy
deformation problems \cite{Kyo78}. We present another proof based on
form factors.
Our proof relies on comparing the generating functions
$\CC_\pm^\infty(z_s)$ with a $N\to\infty$ limit of the $N$-soliton
$\tau$-functions of the affine sinh-Gordon model.
Therefore, we need to do a small detour into the affine sinh-Gordon
model. See ref. \cite{BB92}  for a recent study of this model.
It is a Toda model over the affine Lie algebra $\hat{sl(2)}$.
It involves two fields, which we denote by $\phi$ and $\xi$,
whose equations of motion are~:
\debut
\d_\nu\d_\nu\ \phi &=& 8M^2\sinh(2\phi) \non\\
\d_\nu\d_\nu\ \xi &=& 8M^2 \cosh(2\phi) \non
\fin
Here $M$ in the classical sinh-Gordon mass.
The vacuum solution to these equations is $\phi_{vac}=0$ and
$\xi_{vac}= 2M^2 r^2$ with $r$ the radial distance.
Substracting the vacuum solution, we have~:
\debut
\d_\nu\d_\nu\ \phi &=& 8M^2\sinh(2\phi) \non\\
\d_\nu\d_\nu\ \bar \xi &=&  8M^2\Big[\cosh(2\phi)-1\Big] \label{Va}
\fin
where $\bar \xi =\xi-\xi_{vac}$. Once the field $\phi$ is known,
$\bar \xi$ is computed from it, if the boundary conditions have been
specified.

The affine sinh-Gordon model possesses two $\tau$-functions
$\tau_\pm$ which are defined by~:
\debut
\tau_+\tau_- &=& \exp \(-\bar \xi \) \non\\
\frac{\tau_+}{\tau_-} &=& \exp \(- \phi\) \label{Vb}
\fin
The $\tau$-functions $\tau_\pm^{(N)}$ for the $N$-soliton solutions are
given by
(See \cite{SCMc73} and references therein)~:
\debut
\tau^{(N)}_\pm  = 1+\sum_{p=1}^N (\pm )^p
\sum_{k_1<k_2 < \cdots< k_p}
X_{k_1} \cdots X_{k_p} \prod_{k_i < k_j} \left( {\mu_{k_i} -\mu_{k_j}
\over \mu_{k_i} + \mu_{k_j}} \right)^2 \label{tau}
\fin
with
\debut
X_i = a_i \exp\(2M\sum_s z_s \mu_i^s \) \non
\fin
Here $a_i$ and $\mu_i$ are the parameters of the N-soliton solutions.
The variables $z_s$ correspond to all the commuting flows of the
affine sinh-Gordon hierarchy.
The $\tau$-functions $\tau^{(N)}_\pm$ are solutions of the
quadratic differential equations :
\begin{eqnarray}
\tau_\pm\d_{z_+}\d_{z_-}\tau_\pm - \d_{z_+}\tau_\pm \d_{z_-}\tau_\pm
= - M^2 \tau_\mp^2 \label{Hiro}
\end{eqnarray}
%Here, $z_\pm=x\pm t$ are the light cone coordinates.
These equations, which are the bilinear Hirota form of the affine
sinh-Gordon model, are equivalent to the eqs. (\ref{Va}) for the fields
$\phi$
and $\bar \xi$.

The claim is that $\CC^\infty_\pm(z_s)$ is a $N\to\infty$ limit of
the functions $\tau_\pm^{(N)}$.
The idea is to approximate the
integrals appearing in the definition of $\CC_\pm^\infty$ by
Riemann sums. Therefore, let us set,
\debut
M=\frac{m}{4}\quad;\quad
\mu_i =- \exp(\th_i) \quad;\quad a_i=\frac{\La}{2\pi N} \non
\fin
with $\La$ an arbitrary constant. Taking the limit $N\to\infty$,
$\La$ fixed, gives~:
\debut
\tau^{(N)}_\pm \ \to\
\sum_{n=0}^\infty\ \frac{(\pm)^n}{n!}
\int_{-\La}^{+\La}\
\prod_{j=1}^n \({\frac{d\th_j}{2\pi} X(\th_j|z_s) }\)\
\prod_{i<j} \tanh^2\({\frac{\th_i-\th_j}{2} }\) \non
\fin
Finally, taking the limit $\La\to\infty$ we exactly get the formula
(\ref{IVe}).
Since the $\tau$-functions are solutions of the Hirota
equations (\ref{Hiro}) for any $N$, $a_i$ and $\mu_i$, it follows that
$\CC_\pm^\infty$ is also a solution. Setting~:
\debut
\CC^\infty_\pm(z_s) = \exp\half (-\bar \xi \mp \phi)\non
\fin
then $\phi$ and $\bar \xi$ satisfy eqs. (\ref{Va}).
Integrating these equations and taking into account the fact that
$\phi$ and $\bar \xi$ only depend on the radial distance proves
the result of McCoy et al with the identification $\chi =-\phi$.
The scaling variable $s$ is defined as $s=\frac{mr}{2}$.

Notice that, as it becomes to be familiar
in quantum integrable models \cite{Its90},
the quantum correlation functions are the $\tau$-functions
for a hierarchy of classical integrable non-linear differential
equations.

\section{Miscellaneous remarks.}

$\bullet$ {\it Relation with isomonodromy deformation problems.}

In \cite{Kyo78},
the Ising correlation functions were computed by using the Dirac
equation
for the fermions in an isomonodromy deformation problem.
The fact that isomonodromy
deformations are relevent to the computation of correlation functions
can be
understood from simple physical arguments. Indeed, monodromies of
correlation
functions reflect, and are actually equivalent to, the equal-time
braiding
relations among the fields. But the braiding relations are scale
invariant and
therefore renormalization group invariant. In other words, the
renormalization
group transformations are isomonodromy deformations.

Where do appear the isomonodromy deformations in our approach?
We started from the algebraic equations defining the form factors,
we solved them and then computed the correlations using the exact
expression of the form factors. The point, making the relation with the
isomonodromy deformations apparent, is that the form factor axioms
depend
both on the matrices encoding the braiding relations and on the
$S$-matrix
which also codes the braiding relations since it is given by the
universal
$R$-matrix. Moreover, the mass scale does not show up explicitly in
the form factor axioms. Therefore, computing the form factors is just
an algebraic way to solve an iso-braiding deformation problem.

\bigskip
\noindent $\bullet$ {\it Generalizations.}

Let us describe a way to generalize our approach to other situations.
The proof that the correlation functions satisfy the affine sinh-Gordon
equations relied on their identification with soliton $\tau$-functions
of this hierarchy. This identification is simply the remark that
the $\tau$-functions and the form factors are built from the same
kernel $K(\th_i,\th_j)=\tanh(\th_i-\th_j/2)$. In the context of
$\tau$-functions, this kernel is the expectation value of vertex
operators from which the hierarchy can be reconstructed.

This suggests the following procedure. First, assuming that the form
factors have been determined (e.g. using algebraic methods inherited
from quantum groups), find vertex operator for which they are the
expectation
values,
\begin{eqnarray}
F^{(n)}(\th_1,\cdots,\th_n)^2=\vev{V(\th_1)\cdots V(\th_n)} \non
\end{eqnarray}
Here the expectation values are in an auxiliary Hilbert space on which
the vertex operators are acting. Then, reconstruct the hierarchy
from the knowledge of the vertex operators
(e.g.  using Casimirs in the vertex operator algebra).
The correlation functions will then be soliton-type $\tau$-functions
of the corresponding hierarchy.
The simplest models in this approach are those with diagonal
$S$-matrices
as we will describe elsewhere.

\section{Appendix: Fredholm determinant.}

Here, we show that the correlation functions can be written as Fredholm
determinants. Indeed, using the identity,
\begin{eqnarray}
\prod_{i<j}\left({\frac{\mu_i-\mu_j}{\mu_i+\mu_j}}\right)^2 =
\det\({\frac{2\sqrt{\mu_i\mu_j}}{\mu_i+\mu_j} }\)_{i,j=1,\cdots,n} \non
\end{eqnarray}
the $\tau$-functions $\CC^\infty_\pm(z_s)$ becomes~:
\begin{eqnarray}
 \CC^\infty_\pm(z_s) = \sum_{n=0}^\infty \frac{(\pm)^n}{n!}
 \int^{+\infty}_{-\infty}
\prod_{i=1}^n\({\frac{d\th_i}{2\pi} X(\th_i|z_s)}\)\
\det\({\inv{\cosh(\frac{\th_i-\th_j}{2})}}\)_{i,j=1,\cdots,n} \non
\end{eqnarray}
By definition, this is a Fredholm determinant,
\begin{eqnarray}
\CC^\infty_\pm(z_s) = {\rm Det}\({ 1 \pm K }\) \non
\end{eqnarray}
for an integral operator with kernel,
\begin{eqnarray}
K(\th,\th') = \sqrt{\frac{X(\th|z_s)}{2\pi}}\
\inv{ \cosh(\frac{\th-\th'}{2})}\
\sqrt{\frac{X(\th'|z_s)}{2\pi}} \non
\end{eqnarray}
It is interesting to note that all the dependence in the parameters
$z_s$,
as well as the mass scale, is concentrated in the potential
$X(\th|z_s)$;
i.e. the heart of the kernel, $\cosh^{-1}((\th-\th')/2)$, is
independent of
all these parameters. The property, that all the coupling constants
and all the renormalization group dependence manifest themselves only
in the potential part of the kernel, appears frequently when dealing
with Fredholm determinant in physical context, see e.g. \cite{Its90}.

Finally, notice also that since the square of the Ising model is
equivalent to the sine-Gordon model at the free field point, this
implies that the correlation functions of the sine-Gordon theory
at that particular point can be written as the square of Fredholm
determinants.

\end{document}